\documentclass[preprint,showpacs,showkeys,aps]{revtex4}

\usepackage{graphicx}% Include figure files

\begin{document}

\title{Free Energy Changes, Fluctuations, and Path Probabilities}
\author{
Wm. G. Hoover and
Carol G. Hoover                                              \\
Ruby Valley Research Institute                               \\
Ruby Valley, Nevada 89833                                    \\
}

\date{\today}



\begin{abstract}

We illustrate some of the static and dynamic relations discovered by
Cohen, Crooks, Evans, Jarzynski, Kirkwood, Morriss, Searles, and Zwanzig.
These relations link
nonequilibrium processes to equilibrium isothermal free energy changes
and to dynamical path probabilities.  We include ideas suggested by
Dellago, Geissler, Oberhofer, and Sch\"oll-Paschinger.  Our treatment is
intended to be pedagogical, for use  in an updated version of our book
on {\it Time Reversibility, Computer Simulation, and Chaos}.  Comments
are very welcome.

\end{abstract}

\maketitle

\section{Introduction}

The past 18 years have seen an explosion of activity linking {\em nonequilibrium}
processes to entropy production, fluctuations, and free-energy changes.  This
work connects equilibrium statistical mechanics to far-from-equilibrium
dynamics.  Typically initial conditions are taken from Gibbs' equilibrium
ensembles with the subsequent dynamics featuring explicitly time-dependent
forces.  In the end time averages with the time-dependent forces are related
to equilibrium phase-space averages. This work makes contact with nonequilibrium
steady state analysis as a longtime lowfield limiting case.  Because it sets
the stage for further theoretical advances and practical applications, these
novel force-control activities deserve critical consideration.  We consider
two aspects of this recent research here:
(1) calculating isothermal free energy differences in terms of
far-from-equilibrium work; and
(2) calculating the relative probabilities of {\it nonequilibrium} forward and
reversed trajectories.
We illustrate both these ideas here for a simple
nonequilibrium oscillator problem.

In both cases we can make definite verifiable statements concerning
{\it nonequilibrium} trajectories, by relating these trajectories to
phase-space distributions taken from Gibbs' statistical mechanics. The
new work goes well beyond Green and Kubo's linear-response theory in that
it describes evolving systems far from equilibrium.

In Section II we describe Jarzynski's recent generalization of Kirkwood and
Zwanzig's statistical mechanical perturbation theories.  In Section III we
illustrate these free-energy ideas for a toy-model oscillator, using both Hamiltonian
and Nos\'e-Hoover dynamics.  In Section IV we describe one of Crook's relations
for the relative probabilities of forward and backward trajectories, and apply
it to the same model oscillator system.  Section V
is a summary, putting this work in perspective with other contemporary
efforts to enhance our understanding of nonequilibrium systems.

\section{Fluctuations in Nonequilibrium Steady States}

Gibbs'\cite{b1} (and Boltzmann's) statistical mechanics is the
well-established last word in relating equilibrium thermodynamic properties
to microscopic phase-space states.  {\it Nonequilibrium} properties and processes
lack such a clear foundation.  Their understanding is still in a state of
flux.  Close to equilibrium perturbation theory can be applied.  Green and Kubo
used linear-response theory to treat the linear transport laws---Fick's
diffusion, Newton's viscosity, Fourier's heat conduction. Green and Kubo
expressed the transport coefficients in terms of equilibrium (Gibbsian) time
correlation functions.

{\it Strongly} nonequilibrium situations are still hard to treat
theoretically.  These harder problems include not just the traditional
``approach to equilibrium'' based on Hamiltonian mechanics
and Liouville's theorem, but include also analyses of nonequilibrium steady
states based on deterministic, but mostly nonHamiltonian, thermostats.  These
thermostats can generate fractal steady-state phase-space distribution
functions quite different to Gibbs' smooth canonical distributions\cite{b2}.

Before computer simulation became commonplace perturbation theory was the
most promising basis for a thermodynamic treatment of nonequilibrium
situations.  Kirkwood\cite{b3} and Zwanzig\cite{b4} computed isothermal
free-energy differences directly from Gibbs' equilibrium phase-space
distributions. Here we illustrate their work, as well as the recent
improvements which have extended it.

Kirkwood introduced a ``coupling parameter'', $0<\lambda<1$, used to change
the Hamiltonian by an amount $\Delta {\cal H}$:
$$
{\cal H}(q,p,\lambda) = {\cal H}_0(q,p) + \lambda \Delta {\cal H}(q) \ .
$$
Increasing the coupling parameter from 0 to 1 converts the ``reference
Hamiltonian'' to a new one,
${\cal H}_1(q,p) ={\cal H}_0(q,p) + \Delta {\cal H}(q)  $.
The coupling parameter can be used to ``turn on'' an external field, like
gravity, or to make a change in the interparticle forcelaw.

The corresponding $\lambda$-dependent canonical partition function, $Z$,
and Helmholtz' free energy, $A$,
$$
e^{-A(N,V,T,\lambda)/kT} = Z(N,V,T,\lambda) = \prod
{\textstyle [\int\int dqdp/h]} e^{-{\cal H}(q,p,\lambda)/kT} \ ,
$$
vary smoothly with $\lambda $ from their reference values to the
fully-perturbed ones:
$$
A(N,V,T,1) - A(N,V,T,0) = 
kT\ln\frac{Z(N,V,T,0)}{Z(N,V,T,1)}
\equiv \int_0^1\langle \Delta H \rangle _\lambda d\lambda \ .
$$
The overall response of the energy to the perturbation
$\lambda \Delta {\cal H}$ made it possible to compute the corresponding
(Helmholtz) free-energy change.

19 years later Zwanzig suggested making a ``sudden'', rather than continuous,
change in $\lambda$, providing a simpler formulation of the same free-energy
change:
$$
A(N,V,T,1) - A(N,V,T,0) = 
kT\ln\frac{Z(N,V,T,0)}{Z(N,V,T,1)} =
\langle \Delta {\cal H} \rangle_{\lambda  = 0} \ ,
$$
with the average value of the perturbation computed using the phase-space
distribution of the unperturbed Hamiltonian.  If the perturbed and
unperturbed phase-space distributions are not too different Zwanzig's
approach is the simpler and cheaper of the two procedures.

Kirkwood and Zwanzig's procedures can equally-well be expressed in terms of the
equilibrium phase-space distribution function $f(q,p,T,\lambda )$:
$$
kT\ln[Z(0)/Z(1)]_{\rm Kirkwood} = \int_0^1d\lambda \textstyle{ \prod(\int \int
dqdp)}   f(q,p,T,\lambda)    \Delta {\cal H}(q) \ .
$$
$$
kT\ln[Z(0)/Z(1)]_{\rm Zwanzig} = \textstyle{ \prod(\int \int
dqdp)}   f(q,p,T,\lambda= 0)    \Delta {\cal H}(q) \ .
$$
Both these perturbation-theory approaches assume constant temperature throughout.
Isothermal Monte Carlo simulations are well-suited to these equilibrium
perturbation approaches\cite{b5}.

{\it Dynamical}
isothermal simulations became possible with the isokinetic approach to
thermostating of the 1970s\cite{b6} and especially with the isothermal canonical
thermostat introduced by Nos\'e in 1984\cite{b7,b8}.  With these new 
temperature-control tools it became possible to generalize Kirkwood and
Zwanzig's equilibrium phase-space ideas to far-from-equilibrium dynamical
simulations\cite{b9,b10,b11}.

In 1993 Evans, Cohen, and Morriss suggested a nonequilibrium phase-space
measure applicable to deterministic thermostated steady-state
simulations\cite{b9}.  Evans and Searles clarified the applicability of this
nonequilibrium measure (it becomes correct at long times) the following
year\cite{b10}.

Jarzynski's profound 1997 insight was that the phase-space distribution function
$f(q,p,\lambda)$ need not be an equilibrium quasistatic one so long as the
$\lambda$-dependent trajectory, with $\lambda = \lambda(t)$, follows
Hamiltonian or Nos\'e-Hoover mechanics\cite{b11}.  Then the mechanical
description, through Liouville's Theorem, describes the nonequilibrium flow
of phase-space probablity density.

Let us consider first the familiar Hamiltonian case.  Then the probability
density $f(q,p,\lambda(t),t)$ propagates unchanged, as does also the comoving
phase volume $\otimes(q,p,t)$, even if the underlying Hamiltonian
${\cal H}(q,p,\lambda(t))$ is wildly time-dependent. This means that the
initial equilibrium probability density at
$(q_0,p_0)$: $f(q_0,p_0,\lambda = 0)$ is {\it identical to} the final one,
at $(q_1,p_1)$: $f(q_1,p_1,\lambda = 1)$---provided that the intervening 
$(0 \rightarrow 1)$ dynamics is all Hamiltonian. {\it During} the process the
``work'' done by $\lambda$ along a particular phase-space trajectory is
$W \equiv \int_0^1\Delta {\cal H}d\lambda$.
Consider then the phase-space average, over the unperturbed distribution of
trajectory starting points, of
$e^{-\Delta {\cal H}/kT} = e^{-W/kT}$.  Here the total work $W$ is evaluated
{\it at} the time when $\lambda$, after its specified variation along the
path, has reached its final value of unity:
$$
\textstyle{
\int\int f(q_0,p_0,\lambda = 0)e^{-\Delta {\cal H}(\lambda = 1)/kT}dqdp
 \equiv \int\int f(q_1,p_1,\lambda = 1)dqdp} \longrightarrow 
$$
$$
Z(\lambda = 0) \langle e^{-\Delta H/kT} \rangle _0 = Z(\lambda = 1)
\Longleftrightarrow 
e^{(A_0 - A_1)/kT} \equiv \langle e^{-W/kT} \rangle_0 \ .
$$
This last equation, Jarzynski's remarkable free-energy identity (or
``equality'', or ``relation'', or ``Theorem'') gives the
{\it equilibrium} free-energy difference in terms of arbitrarily
{\it far-from-equilibrium} phase-space trajectories.

The dynamics need not be Hamiltonian.  Jarzynski extended the free-energy
relation to include thermostated Nos\'e-Hoover dynamics\cite{b11}.  In
that thermostated case the comoving phase volume changes with time:
$$
\frac{\otimes(t)}{\otimes(0)} = \frac{f(0)}{f(t)} =
\frac{(dqdpd\zeta)_t}{(dqdpd\zeta)_0} \equiv
 e^{-\int_0^t\zeta(t^\prime)dt} \ ,
$$
where $\zeta(t^\prime)$ is the Nos\'e-Hoover thermostating friction
coefficient.  In the end the free-energy result is exactly the same, as
we will detail and check numerically in the next Section, with a
simple example problem.

Naturally enough, some doubted these surprising results.  See, for
instance, references [12] and [13].  After some prodding by naysayers,
Jarzynski published more details and extensions of his proofs, valid for
stochastic thermostats as well as for the Hamiltonian and Nos\'e-Hoover
situations, and including {\it any} time-reversible equations of motion which
maintain the canonical distribution\cite{b13}.  Let us consider the
Nos\'e-Hoover case in more detail.

For the Nos\'e-Hoover motion equations an ``extended'' (to include $\zeta$)
canonical distribution is maintained by the equations of motion:
$$
\{ \ \dot q = p/m \ ; \ \dot p = F - \zeta p \ \} \ ; \
\dot \zeta = [(K/K_0) - 1]/\tau^2 \Longleftrightarrow
f \propto e^{-{\cal H}/kT}e^{-\zeta^2\tau^2/2} \ .
$$
Here $\zeta$ is a control variable, controlling the kinetic energy
$K \propto p^2$. $K_0$ is the kinetic energy corresponding to the specified
temperature $T$
and $\tau$ is the (arbitrary) relaxation time controlling the timescale
of the resulting kinetic temperature fluctuations.
This same derivation can be carried through with the generalized
distribution  [including the factor $e^{-(\zeta \tau)^2/2}$] at time 0.
Multiplying that distribution by the exponential of $-W/kT$, gives again
the generalized distribution when $\lambda$ reaches unity and establishes
Jarzynski's free-energy relation for Nos\'e-Hoover dynamics.  The only new
aspect is the need to take the changing phase-volume change into account, as is
most clearly described by Sch\"oll-Paschinger and Dellago\cite{b14}.  Let us
illustrate the Kirkwood-Zwanzig-Jarzynski ideas with a simple example, a
harmonic oscillator whose force constant varies with time.

\section{The Free Energy Theorems Illustrated for an Oscillator}

To illustrate first Kirkwood's coupling-parameter idea, as extended later
by Zwanzig and Jarzynski, consider a thermostated (canonical distribution)
one-dimensional harmonic oscillator with unit
mass.  We choose to increase the oscillator force constant $\kappa$
from unity to four as $\lambda(t)$ is slowly increased (``quasistatically'';
``reversibly'') from 0 to 1:
$$
1 < [\kappa \equiv 1 + 3\lambda] < 4 \longrightarrow
$$
$$
Z(\lambda) = (1/h)\int_{-\infty}^{+\infty} e^{-p^2/2kT}dp
\int_{-\infty}^{+\infty} e^{-(1+3\lambda)q^2/2kT}dq =
kT/h\nu \ ; \
2\pi \nu = \sqrt{1+3\lambda} \ .
$$
$$
\phi = (1/2)q^2[1+3\lambda] \rightarrow
\ln\frac{Z(N,V,T,0)}{Z(N,V,T,1)} =
 \ln\left[\frac{(kT/h\nu_0)}{(kT/h\nu_1)}\right] = 
 \ln\left[\frac{(kT/h\nu_0)}{(kT/2h\nu_0)}\right] = \ln2  \ .
$$
For simplicity we choose unit temperature, $kT = \langle p^2/m \rangle = 1$.
The completed perturbation reduces the rms fluctuations in $\langle q^2 \rangle$
by a factor of two and increases the Helmholtz free energy by $kT\ln 2$.
Kirkwood's quasistatic analysis calculates the partition-function
ratio by integration of the work associated with the perturbation:
$$
\ln\frac{Z(N,V,T,0)}{Z(N,V,T,1)} =
-\int_0^1(\partial \ln Z/\partial \lambda )d\lambda =
\int_0^1(3/2)d\lambda/(1+3\lambda) = (1/2)[\ln(4)-\ln(1)] = \ln2 \ .
$$
From this standpoint the free energy difference is simply the integrated work
needed to make a very gradual, reversible change in the Hamiltonian.

Zwanzig's approach is standard perturbation theory.  It corresponds
physically to a {\it sudden} change in the Hamiltonian,
$$
{\cal H} = (q^2 + p^2)/2 \stackrel{\rm sudden}{\longrightarrow}(4q^2 +
p^2)/2 \ .
$$
The free energy change can then be written as the average value of the
perturbation's effect on the unperturbed ($\lambda = 0$) canonical
probability density:
$$
A(N,V,T,0) - A(N,V,T,1) = 
kT\ln \langle e^{-(3/2)q^2/kT}\rangle_0 = kT\ln(kT/4)/(kT/2) = - kT\ln(2) \ .
$$

Jarzynski's nonequilibrium generalizations of these Gibbsian ideas can be
checked by using either Hamiltonian or Nos\'e-Hoover dynamics to implement
the perturbation.  This choice is not entirely coincidental.  Dettmann
showed that single-temperature Nos\'e-Hoover mechanics has its roots in an
underlying Hamiltonian\cite{b15}.

Using a relaxation time of unity, and unit temperature, the equations of
motion for
$(0 < t < 1)$ in the two cases are:
$$
\dot q = p \ ; \ \dot p = -q(1 + 3t) \ ; \
\dot I = 3q^2/2 \  {\rm [Hamiltonian]} \ ;
$$
$$
\dot q = p \ ; \ \dot p = -q(1 + 3t) - \zeta p \ ; \ \dot \zeta = p^2 - 1 \ ; \
\dot I = 3q^2/2 \ {\rm [Nos\acute{e}-Hoover]} \ .
$$
For each trajectory we compute the exponential of the work integral 
$$
e^{-\int_0^1W(t^\prime)dt/kT} = e^{-W/kT} = e^{I_0 - I_1} \ .
$$
The complete trajectory integral, $I(1) - I(0)$ includes the work
done on the oscillator by the perturbation, and includes also (in the
Nos\'e-Hoover case) the changing phase volume.
$$
\int_0^1\dot Idt = W/kT \ {\rm [Hamiltonian]} \ ; \
\int_0^1\dot Idt = W/kT\ln[\otimes(0)/\otimes(1)] \
{\rm [Nos\acute{e}-Hoover]} \ .
$$                 
In the Nos\'e-Hoover case this last result follows from the useful identity:
$$
\dot {\cal H}_\lambda =
(d/dt)[(q^2(1+3t) + p^2 + \zeta^2)/2] = (3q^2/2) - \zeta \Longrightarrow
$$
$$
\textstyle{
\int\int\int(dqdpd\zeta/h)_0 e^{-{\cal H}_0/kT}e^{-W/kT} =
\int\int\int(dqdpd\zeta/h)_1 e^{-[4q^2+p^2+\zeta^2)/2]} \equiv Z_1 \ .
}
$$
The expression in square brackets is the ``extended'' Nos\'e-Hoover
Hamiltonian ${\cal H}_\lambda$.  In the Nos\'e-Hoover case the
``extended'' partition functions $Z_0$ and $Z_1$ include integrations
over the thermostat variable $\zeta $.

For numerical Runge-Kutta integration of trajectories we chose
the initial values of $(q,p)$ or $(q,p,\zeta)$ from equilibrium (Gaussian)
distributions $\propto e^{(-q^2 - p^2)/2}$ or $e^{(-q^2 - p^2 - \zeta^2)/2}$.
We verified Jarzynski's relations numerically by choosing and analyzing
{\it millions} of trajectories.  We computed average values of the left side,
$\langle e^{-W}\rangle$,
and the right side, $e^{\Delta A/kT}$, of the Jarzynski relations.
For 100 million collisions, with either Hamiltonian or Nos\'e-Hoover mechanics,
his predicted average value of the partition-function ratio, $I =(1/2)$, is
nicely reproduced, with four- to five-figure accuracy, perfectly consistent
with both versions of Jarzynski's 1997 theorem.  Such Theorems are
also valid for other deterministic and stochastic equations of motion as is
detailed in References [16]-[19].

\section{Crook's Reversibility Relation}

Crooks and Jarzynski used a variation of these ideas\cite{b16} to compute
the relative probabilities of forward and reversed nonequilibrium
trajectories.  Their work makes contact with the seminal 1993 and 1994
papers of Evans, Cohen, Morriss, and Searles\cite{b9,b10}.  Though Crooks
and Jarzynski emphasize stochastic dynamics we illustrate one of their
results here for time-reversible deterministic Hamiltonian dynamics.  This
version of their Theorem (or ``result'', or ``relation'', or ``identity'')
can also be viewed as an illustration (or consequence) of ``detailed balance'' or
``microscopic reversibility'', the equal likelihood of forward and backward
trajectories linking two microstates.  For convenience we will choose those
as states of the time-dependent oscillator at times 0 and 1.

In order to understand their formulation of the problem let us compare two
probabilities: the ``forward'' orbit of a canonical oscillator going from
the phase point $(q_0,p_0)$ at time 0 to $(q_1,p_1)$ at time 1; the
``backward'' orbit of the oscillator, equilibrated at $(q_1,-p_1)$ and going
backward to $(q_0,-p_0)$.  On the forward orbit the force constant increases
from 1 to 4; on the backward orbit the force constant decreases, from 4 to
1.  The probability for the forward and reversed trajectories are not equal
because the initial energies and free energies differ.  In the forward and
backward directions the normalized rates are proportional to
$$
e^{(A-H)_0/kT}(dqdp)_0 \ ; \ e^{(A-H)_1/kT}(dqdp)_1 \ ,
$$
with
$$
\Delta H = H_1 - H_0 = W = \int_0^1(3/2)q^2dt \ ; \
\Delta A = A_1 - A_0 = kT\ln2 \ .
$$
Evidently the ratio of the forward and backward rates is:
$$
\frac{e^{(A-H)_0/kT}(dqdp)_0}{e^{(A-H)_1/kT}(dqdp)_1} = e^{(-\Delta A + W)/kT} \ .
$$
The $dqdp$ phase-volume elements cancel according to Liouville's Theorem.
This is (one of) Crook's ``Fluctuation Theorems'', a simple consequence of
the time reversibility of the Hamiltonian equations of motion.  Like the
free energy theorems, it can easily be checked by Runge-Kutta integration.

If we view the extra dissipated work as a time-integrated entropy production,
$$
\Delta S = (W - \Delta A)/T \ ,
$$
then this fluctuation theorem has the same form as the longtime steady-state
relation\cite{b9}
$$
\ln\left[\frac{prob(+\sigma)_\tau}{prob(-\sigma)_\tau}\right] = [\sigma \tau/k] \ ,
$$
where the entropy production rates $\sigma$ are observed average values during an
averaging ``time window'' of length $\tau$.

\section{Perspectives}

Jarzynski's isothermal free-energy relations link equilibrium statistical
mechanics to nonequilibrium dynamics.  Though the Jarzynski work includes
arbitrary time-dependent
forces this approach requires equilibrium initial conditions and is limited
to a single fixed temperature $T$.  We still need theories for treating
more general nonlinear problems, such as stationary shockwaves.  Shockwaves
link together equilibrium states with two different temperatures.  Within
the shockwave the temperature is a tensor, with
$T_{xx} \neq T_{yy}$\cite{b18}.

We
still need good methods for understanding the
fractal phase-space structures of systems with two or more temperatures,
not just one.  So far the only connection between the
smooth Liouville-Theorem approach of the present work and truly
nonlinear steady states is at the linear level of the Central Limit Theorem
and Green-Kubo linear response.

The processes for which the recent theorems are relevant involve entropy changes
of only a few $k$ and energy changes of only a few $kT$.  Macroscopic
consequences of the Fluctuation Theorems are simply unobservable.
Even so, from the theoretical standpoint the new ideas are stimulating, relating
``real'' finite-rate processes to equilibrium thermodynamics.  Unlike
Kirkwood and Zwanzig's earlier efforts, the new approaches do have counterparts
in micron-scale laboratory experiments in which long molecules or latex balls
are manipulated with external forces.  The back-and-forth fluctuations in
these many-body systems can be approximated with theoretical results for
thermostated models with a few degrees of freedom.  See the reviews [19] and
[20] for references.  Evidently all three routes to understanding, theory,
simulation, and experiment are linked
together by these revolutionary ideas.

From the computational standpoint the new theorems are not particularly ``useful''
for free-energy work.  A careful investigation showed that
the original ideas of Kirkwood and Zwanzig are less costly than
Jarzynski's in making numerical free-energy estimates\cite{b17}.

The literature on this subject is particularly hard to follow.  There is a
tension between derivations treating the most general case possible and
special cases, which are easier to grasp.  Fairly widespread confusion
is evident from the several papers  purporting to find flaws in the
published proofs. There is an enduring need for more
example problems to explore the useful ranges of the many theorems.

The derivations make use of the detailed reversibility of Hamiltonian
or thermostated mechanics.  Systems away from equilibrium for a few
Lyapunov times cannot be reversed due to Lyapunov instability.  Typically
the phase volume shrinks, forward in time, so that the reversed trajectory,
with growing phase volume, is dynamically unstable, and soon seeks out a
phase-space attractor rather than following the repellor.  Shockwaves illustrate
another difficulty in reversing irreversible processes.  From the
macroscopic standpoint an attempt to reverse a strong shockwave would result
instead in a nearly isentropic rarefaction fan.  Whether or not this
qualitative irreversibility
will have a counterpart in the highly nonequilibrium systems
which can be treated theoretically remains to be seen.

\end{document}